# The Biological Big Bang: The First Oceans of Primordial Planets at 2-8 Million Years Explain Hoyle/Wickramasinghe Cometary Panspermia

Carl H. Gibson[*]

UCSD, La Jolla, CA, 92093-0411, USA

**ABSTRACT**

Hydrogravitional-dynamics (HGD) cosmology of Gibson/Schild 1996 predicts that the primordial H-He$^4$ gas of big bang nucleosynthesis became proto-globular-star-cluster clumps of Earth-mass planets at 300 Kyr. The first stars formed from mergers of these 3000 K gas planets. Chemicals C, N, O, Fe etc. created by stars and supernovae then seeded many of the reducing hydrogen gas planets with oxides to give them hot water oceans with metallic iron-nickel cores. Water oceans at critical temperature 647 K then hosted the first organic chemistry and the first life, distributed to the $10^{80}$ planets of the cosmological big bang by comets produced by the new (HGD) planet-merger star formation mechanism. The biological big bang scenario occurs between 2 Myr when liquid oceans condensed and 8 Myr when they froze. HGD cosmology explains, very naturally, the Hoyle/Wickramasinghe concept of cometary panspermia by giving a vast, hot, nourishing, cosmological primordial soup for abiogenesis, and the means for transmitting the resulting life forms and their evolving chemical mechanisms widely throughout the universe. A primordial astrophysical basis is provided for astrobiology by HGD cosmology. Concordance ΛCDMHC cosmology is rendered obsolete by the observation of complex life on Earth.

## 1. INTRODUCTION

When photons decouple from plasma 300,000 years after the big bang, the phase change triggered gravitational formation of Earth-mass gas planets in million-solar-mass clumps as predicted by Gibson (1996)[1] and observed independently by Schild (1996)[2]. Plasma protogalaxies fragmented entirely into such clumps of planets, now frozen to form the dark matter of galaxies and the primary custodians of evolving life. A common assumption is that life originated, and is probably confined, to the planet Earth, and that life has no astrophysical significance. Neither assumption appears to be correct. Over many decades, Hoyle and Wickramasinghe (1977[3], 1982[4], 2000[5]) pioneered the concept that cometary-panspermia explains life on Earth, citing strong spectral evidence that polycyclic-aromatic-hydrocarbons PAH are biological in origin, and that the close morphological similarity between microfossils found in carbonaceous meteors with their terrestrial counterparts is proof of cosmic biological dispersal. Hydro-Gravitational-Dynamics HGD cosmology strongly supports cometary-panspermia as an active component of cosmic and biological evolution from the beginnings of the gas epoch soon after the big bang, Gibson, Schild & Wickramasinghe (2010)[6] until today. Table 1 summarizes the timing of gravitational structure formations from HGD with the indicated time of formation of living organisms. The standard cosmological model leaves no possibility for the chemically complex life observed on Earth to form or be transmitted on cosmological scales. The first stars do not appear until after 400 million years of dark ages along with a handful of planets per star, not the 30 million per star expected from HGD.

---

[*] Corresponding author: Depts. of MAE and SIO, CASS, *cgibson@ucsd.edu*, **http://sdcc3.ucsd.edu/~ir118**

Table 1. HGD structures formed in the plasma and gas epochs as a function of time after the big bang event. Because the universe is expanding, gravitational instability favors density minima. Structure formation is therefore by fragmentation as the minima form voids.

| | |
|---|---|
| Protosuperclusters and protosupercluster voids are formed under photon viscosity control | 30,000 years ($10^{12}$ s), plasma fragments at globular star cluster density, $\rho = 10^{-17}$ kg m$^{-3}$ |
| Protogalaxies fragmented along vortex lines | 300,000 years ($10^{13}$ s), plasma to gas transition |
| Planets form in clumps, stars form and die from planet mergers, life begins | 330,000 years ($10^{13}$ s), hot hydrogen gas planets form water |
| Intelligent life ~ 2 Myr | ~13.7 Gyr ($4 \times 10^{17}$ s) |

Cometary-panspermia CP and hydrogravitational dynamics HGD cosmology are complimentary concepts that must be merged to explain new observations from new space telescopes and improving ground based telescopes. Each of these concepts is needed to fully explain the other. Without HGD, CP has no life seeds to disperse and no means to disperse them beyond a few planets near a star if, against all odds, life should appear on one of these few planets. With HGD, cometary panspermia is an intrinsic part of star formation from planets and their fragments, and the formation and dispersion of life on cosmic scales, possibly intelligent life, occurs soon after the formation of the first chemicals and hot water oceans at about two million years.

## 2. THEORY

Suppose a large volume of fluid exists at rest with uniform density $\rho$. All the fluid within Hubble radius $L_H = ct$ will begin to move toward the origin $\mathbf{x} = 0$ for time $t > t_0$ if a non-acoustic mass perturbation of size $M' = +\delta\rho L^3$ is introduced at $t = t_0$, and all the fluid will begin to move away from the origin if $M' = -\delta\rho L^3$, where c is the speed of light and L is a radius smaller than the Jeans acoustic scale $L_J$ but larger than the largest Schwarz scale $L_{SXmax}$ (both defined below). It can be shown (Gibson 2000)[7] that an isolated mass perturbation grows or decreases exponentially with time squared according to the expression

$$M'(t) = |M'(t_0)| \exp[\pm 2\pi\rho G t^2]$$

where the gravitational (free fall) time scale $\tau_g = (\rho G)^{-1/2}$. Nothing much happens at the origin before the free fall time $\tau_g$ arrives. For $M'$ positive a sudden increase in density occurs near the "cannonball" origin as mass piles up. Hydrostatic equilibrium is established as local pressure increases to balance gravity. Density and temperature increase and hydrodynamic effects become relevant. For $M'$ negative, density near the origin suddenly decreases and a void appears surrounding the "vacuum beachball" origin at time ~ $\tau_g$ and propagates radially outward as a rarefication wave limited in speed by the speed of sound. Matter near the void falls radially away due to gravity. Thus, in an expanding universe, gravitational structure formation occurs mostly by fragmentation since the expansion of space favors void formation and resists condensation. Formation of voids in the plasma epoch account for the prominent sonic peak in the cosmic microwave background CMB temperature anisotropy spectrum, rather than acoustic oscillations of plasma that falls into gravitational potential wells of condensed CDM seeds, as assumed by the standard model.

The viscous Schwarz scale $L_{SV} = (\gamma\nu/\rho G)^{1/2}$ is the length scale at which viscous forces match gravitational forces, where $\gamma$ is the rate-of-strain, $\nu$ is the kinematic viscosity of the fluid and G is Newton's gravitational constant. The turbulent Schwarz scale $L_{ST} = \varepsilon^{1/2}/(\rho G)^{3/4}$, where $\varepsilon$ is the viscous dissipation rate of turbulent kinetic energy. The diffusive Schwarz scale $L_{SD} = (D^2/\rho G)^{1/4}$, where D is the diffusivity of the fluid.

The Jeans acoustic scale $L_J = V_S \tau_g$. Only the Jeans (1902) criterion that gravitational structure formation cannot occur for scales $L < L_J$ is used in the standard cosmological model. This is its fatal flaw. Jeans neglected the effects of viscosity, turbulence and diffusivity. Because the scale of causal connection ($L_H = ct$) is smaller than the Jeans scale during the plasma epoch, no structure can form. Thus it was necessary to invent cold dark matter CDM; that is, a mythical substance with the property that its Jeans scale $L_J$ is smaller than $L_H$ because it is cold. The standard model assumes CDM is nearly collisionless, like neutrinos, which means its diffusivity D is very large so that its $L_{SD}$ scale during the plasma epoch exceeds $L_H$. Thus, whatever CDM is, it cannot condense during the plasma epoch according to the hydrogravitational dynamics HGD Schwarz scale criteria (Gibson 2009a, b, 2010)[8,9,10].

## 3. STRUCTURE FORMATION IN THE EARLY UNIVERSE

Figure 1 illustrates schematically the differences between HGD cosmology and ΛCDMHC cosmology during the plasma epoch, soon after mass became the dominant cosmological component at time ~ $10^{11}$ seconds after the big bang event over energy. From HGD, 97% of the mass at that time is non-baryonic, with the weakly collisional properties and mass of neutrinos (green). The rest (yellow) is hydrogen-helium plasma (protons, alpha particles and electrons). The total mass is slightly more than required for the universe to be flat, leading to an eventual big crunch. For ΛCDMHC the anti-gravity dark-energy density eventually dominates and accelerates the expansion of an open universe as the kinetic-energy and gravitational-potential energy densities approach zero.

In Fig. 1, CDM seeds (top left) cannot possibly condense because this non-baryonic dark matter NBDM (green) is nearly collisionless and therefore super diffusive. Neither can NBDM hierarchically cluster HC to form potential wells, into which the plasma (yellow) can fall and produce loud acoustic oscillations (top right). Gravitational structure formation begins in HGD (bottom left) as fragmentation at superclustermass scales (~$10^{46}$ kg) to form empty protosuperclustervoids, shown as open circles. Because the universe is expanding, structure formation is triggered when viscous forces permit it at density minima to form voids. Voids expand as rarefaction waves limited by the speed of sound, which in the plasma epoch is $c/3^{1/2}$, where c is the speed of light. Fragmentation continues at smaller and smaller mass scales to that of protogalaxies (~$10^{43}$ kg). Weak turbulence develops at the expanding void boundaries. Turbulence determines the morphology of plasma protogalaxies, which fragment along turbulent vortex lines where the rate of strain γ is maximum. The plasma-turbulence Nomura scale $10^{20}$ m is that of all protogalaxies at their transition to gas[11].

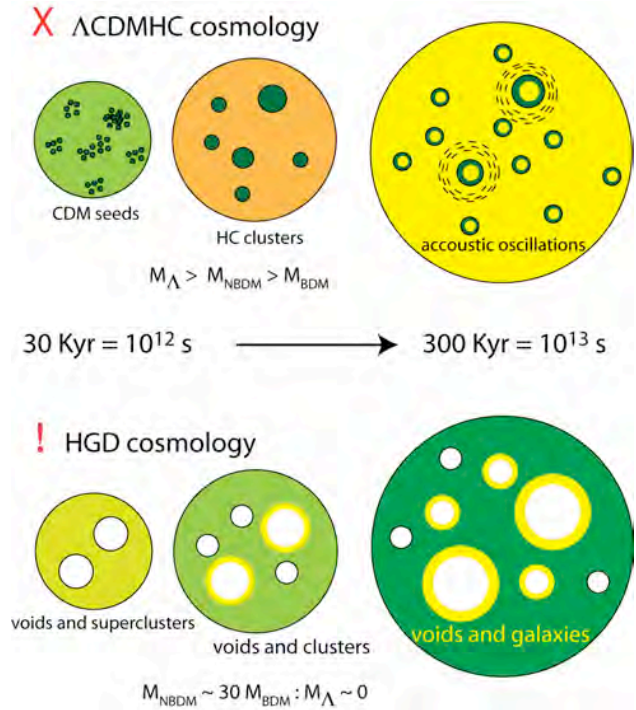

Figure 1. Comparison of plasma epoch behavior of the universe for the incorrect **X** standard ΛCDMHC model (top) versus the correct **!** hydrogravitational dynamics HGD model (bottom). Condensation of the nearly collisionless non-baryonic dark matter to form CDM seeds that cluster (top) is physically impossible. Structure formation occurs by fragmentation (bottom) when the increasing scale of causal connection $L_H = ct$ matches the viscous Schwarz scale $L_{SV}$. This occurs at $t_0 = 10^{12}$ s (30 Kyr). Proto-galaxies produced at plasma-gas transition[11] have Nomura turbulence morphology and length scales reflecting $t_0$.

Figure 2 illustrates schematically the two cosmologies in the gas epoch from 300 Kyr to 300 Myr ($10^{13}$ -$10^{16}$ s), which is often termed the dark ages for ΛCDMHC (top) because this is the time required for the first star and the first planets to appear in this cosmology. The temperature of space has fallen to a few °K that will freeze any gas. The density has decreased by a billion. Any life arising on the handful of planets produced as stars form from gas would likely be blasted out of existence by superstars so powerful they re-ionize all the plasma of the universe as they explode. Extra-terrestrial (and terrestrial) life is virtually impossible by the standard ΛCDMHC model. Life would be extremely rare and confined to local star systems.

For HGD cosmology (Fig. 2 bottom), the $10^{13}$ -$10^{16}$ s interval has many stars and warm planets, and is the optimum time period for life to appear and for its first seeds to be widely scattered on cosmic scales. Plasma-protogalaxies form near the end of the plasma epoch by fragmentation along turbulent vortex lines. Gas-protogalaxies fragment into $10^{36}$ kg Jeans mass clumps of $10^{24}$ kg planetary-mass clouds that shrink and eventually freeze solid as H-$^4$He gas planets (termed primordial-fog-particles PFPs, or micro-brown-dwarfs μBDs) as they cool.

This is the first gravitational condensation. The planet-clump density $\rho_0 = 4\times10^{-17}$ kg m$^{-3}$ is that of the plasma when it first fragmented at $10^{12}$ s. It is no coincidence that this is the observed density of old globular star clusters OGCs. Clumps of planet-mass clouds are termed proto-globular-star-clusters PGCs. No dark age exists for HGD because the first stars appear in a gravitational free fall time $\sim(\rho_0 G)^{-1/2} = 2 \times 10^{13}$ s or less at temperatures too high for it to be dark anywhere. The process of star formation is binary-planet-mergers within these PGC clumps of primordial-fog-particle PFP planets.

In HGD cosmology all stars form within PGC clumps of planets. The empty volume without planets is termed the Oort cavity, with size $(M/\rho_0)^{1/3}$, where M is the mass of the central star or stars produced by planet merging from a mass density of planets $\rho_0$. This is $3 \times 10^{15}$ m for a solar mass, which closely matches the observed distance to the Oort cloud of long period comets. From HGD, planets continue to fall into the Oort cavity from its inner boundary, eventually overfeeding the central star or stars to the point that they explode as supernovae. Supernovae are the source of carbon, oxygen, nitrogen etc. that are needed to produce life. We see that all the conditions for the first life in the big bang universe are achieved soon after the plasma to gas transition. Because many of the planets and their moon and meteorite fragments will be relatively warm, many domains of liquid water will exist in this early period as primordial soup kitchens. Once the first templates for life appear, they will rapidly be transmitted to other planets and their associated venues so that life can evolve by mutation and evolution as we see it on Earth.

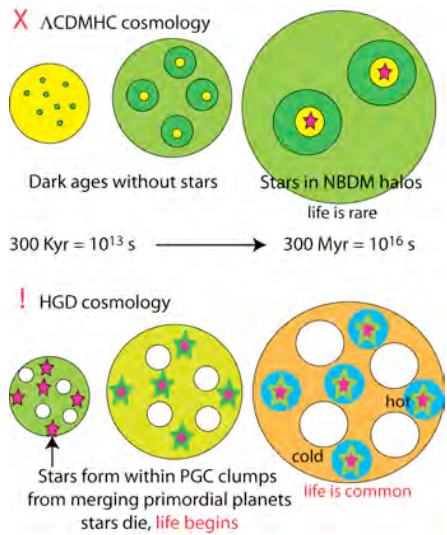

Figure 2. Gas epoch for 300 Kyr to 300 Myr, corresponding to the dark-ages period for ΛCDMHC (top) before the first stars and planets. Astrobiology would be very different in ΛCDMHC because, even if life were to form, it would be confined to a handful of planets formed simultaneously and close to their stellar mother, with no common cosmic ancestors. Life in HGD cosmology (bottom) begins immediately with excellent cosmical mixing within PGC clumps at all stages, some mixing between clumps and within galaxies, and progressively less mixing at larger scales. The non-baryonic dark matter (probably neutrinos) fragments at its LSD scale after the plasma epoch to form galaxy cluster halos (green bottom center). Hot x-ray halos with small mass surround the clusters (blue bottom right). Radio telescope detection (Rudick et al. 2008) of completely empty supervoids with scales $\sim 10^{25}$ m contradict the ΛCDMHC model (top), which makes small voids that are not empty at its last stage rather than its first.

Figure 3 shows optical band images from the Hubble Space Telescope HST of the Helix planetary nebula PNe, one of the closest PNe to the Earth at a distance of $\sim 6 \times 10^{18}$ m[12]. The central white dwarf is constantly fed PFP planets as comets falling in from the Oort cavity boundary, which it converts to carbon. Part of the gas of the comet-planets released to the star is expelled as a plasma beam by the rapidly spinning white dwarf. The plasma beam partially evaporates and photoionizes the nearest surrounding planets to ~ $10^{13}$ m scales (depending on the mass of the frozen planet) so they are detectable. When the white dwarf mass exceeds the Chandrasekhar limit of 1.44 solar mass ($2.2 \times 10^{30}$ kg) it explodes with large predictable brightness, making it an excellent standard candle for distance estimates.

If the line of sight to the Earth intersects an evaporated planet atmosphere it is slightly dimmed, otherwise it is not dimmed (red circles) and no dark energy is needed to explain the dimming. Thus dark energy Λ is a systematic SNeIa dimming error of dying carbon stars embedded in clumps of primordial planets, not 70% of the density of the present universe as claimed by ΛCDMHC cosmology. A similar systematic error in the age of the universe (Sandage et al. 2006) can also be explained as systematic SNeIa dimming errors due to PFP planet atmospheres (bottom left, red circles). Rather than 16 Gyr, the corrected age is 13.7 Gyr, consistent with the value estimated from cosmic microwave background evidence.

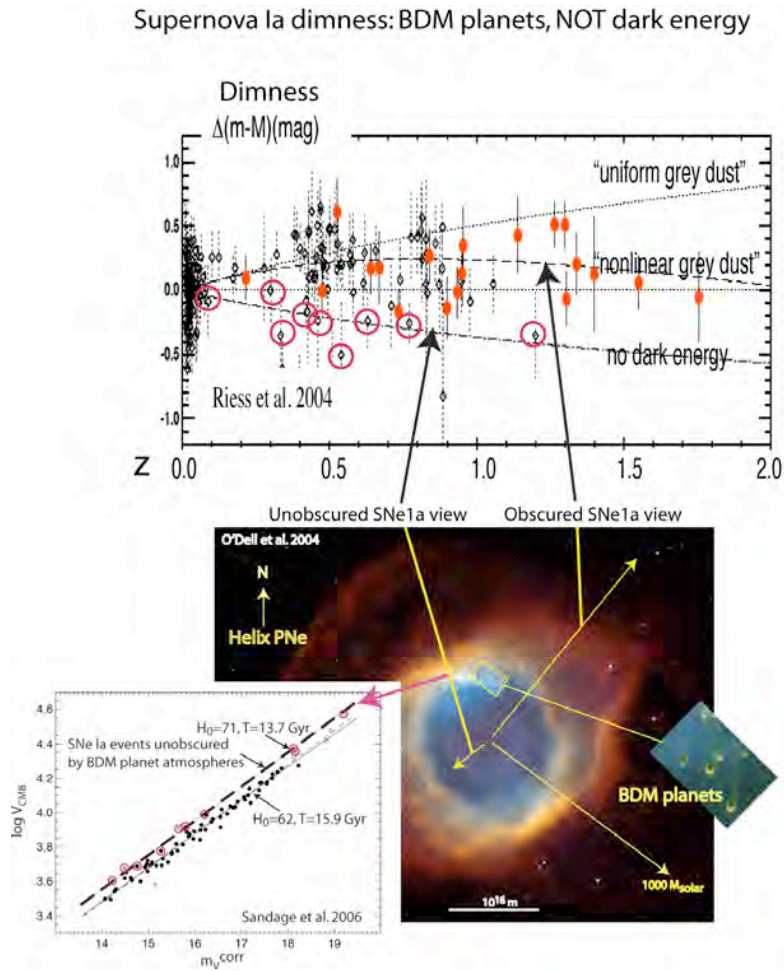

Figure 3. Helix Planetary Nebula (O'Dell et al. 2004) interpreted using HGD cosmology (Gibson 2010, Fig. 8)[10]. At the top is supernova Ia (SNeIa) dimness evidence of dark energy (red ovals) not taking effects of dark matter planets in clumps as the source of all stars and the dark matter of galaxies. Unobscured SNeIa lines of sight (red circles) are brighter, suggesting Λ is a systematic dimming error corrected by HGD cosmology. A similar systematic dimming error due to evaporated PFP atmospheres is suggested at bottom right for the age of the universe (Sandage et al. 2006) inferred from the dimmed SNeIa to be 16 Gyr: corrected (red circles) to 13.7 Gyr. A yellow arrow (bottom right) shows the radius of a sphere containing a thousand solar masses of PFP planets.

Most of the evaporated planets of Helix in Fig. 3 are obscured by polycyclic aromatic hydrocarbon PAH dust. Only 5000 optical planets can be seen. Infrared space telescopes such as Spitzer reveal 40,000 or more, as well as planet protocomets falling into the central star.

## 4. EVIDENCE OF PRIMORDIAL PLANETS FROM INFRARED TELESCOPES

Figure 4 shows the Helix PNe in both optical (bottom) and infrared (top) frequencies (Matsuura et al. 2009)[13]. Numerous planets and all of the planet-protocomets are quite invisible in the HST optical frequencies.

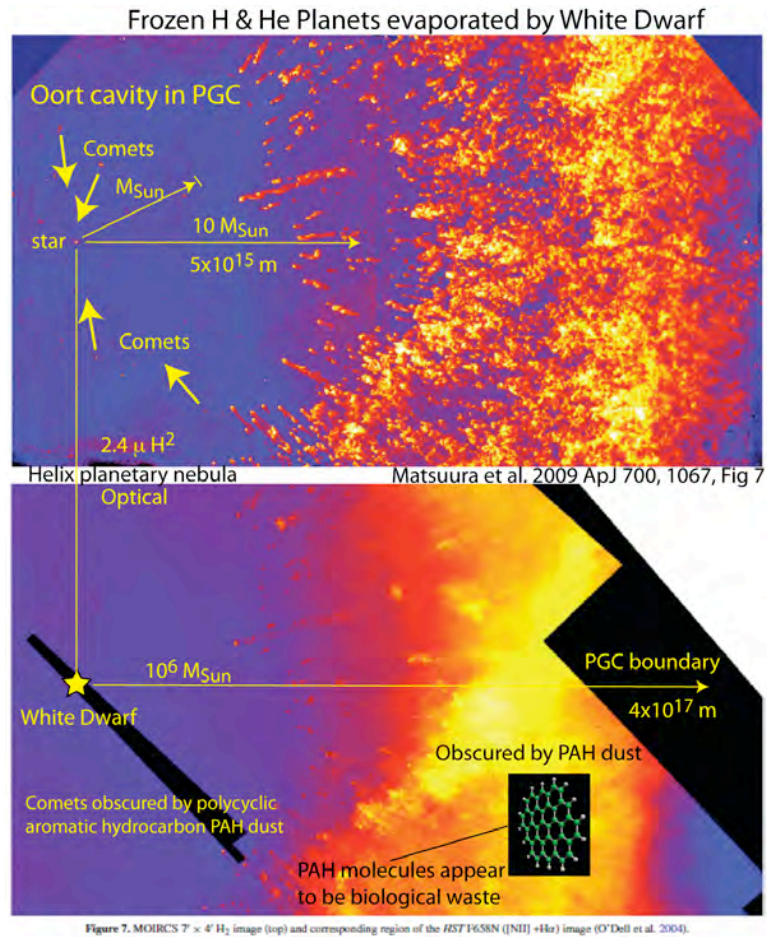

Figure 4. Helix PNe reveals partially evaporated dark matter planets (top) at the hydrogen molecule $H^2$ wavelength 2.4 µ (Matsuura et al. 2009)[13] that are obscured to an extent that increases with radius from the central white dwarf star, suggesting PAH dust has been evaporated from the planets as well as their originally frozen H-He gas. A typical PAH molecule is shown (bottom right), which has the structure of carbohydrate oils used as food by terrestrial organisms.

Note from Fig. 4 that planets at the Oort cavity boundary are sorted according to their amounts of biological activity by the radiation pressure, which increases with PAH content because the heat transfer increases with the size of the atmosphere.

Figure 5 is a similar Helix PNe comparison (Meaburn & Boumis 2010)[14], showing clumps (globulettes)[15] of PFPs revealed in the infrared that are invisible in the optical. Clumping is easy to understand from HGD since the plasma beam radiating from the white dwarf magnetic pole is responsible for the partial evaporation and photoionization of the dark matter frozen gas planets it intersects. A clockwise helical track of evaporation can be traced in Helix, which accounts for its name. The large atmospheres of the irradiated planets causes frictional forces between planets whose atmospheres overlap that may cause them to clump and in some cases merge.

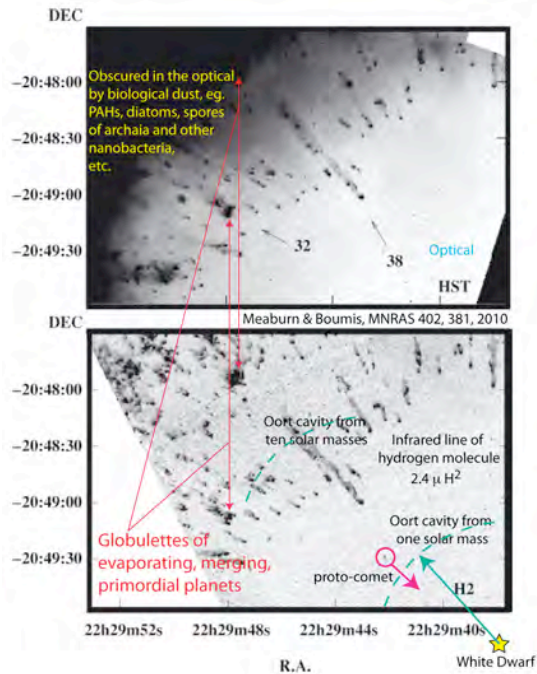

Figure 5. Helix PNe at optical frequencies (top) compared to the infrared molecular hydrogen band (bottom). Globulettes[15] of evaporating, merging primordial planets are inferred from vertical red arrows left of center connecting optical and infrared objects.

Earth is constantly bombarded by meteors. Carbonaceous meteorites such as Murchison contain ample evidence of ancient extra-terrestrial life. As shown in Figure 5, the morphology of fossils contained is that of organisms that were once alive, but most species are unknown or different from those on this planet. The complexity of the organic compounds detected in a sample of the 100 kg object is estimated to exceed a million[16], a significantly larger chemodiversity of its source compared to that of Earth. Evidence suggests Murchison is older than the Earth. Images of cyanobacteria from Muchison and Orgueil meteorites are available at Richard Hoover's website, along with numerous references, http://www.panspermia.org/hoover4.htm.

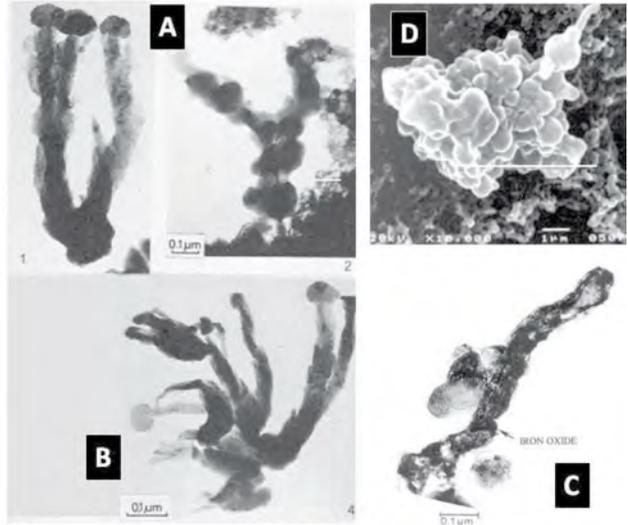

Figure 6. Microbial fossils from comets. Images A, B are bacterial microfossils in the Murchison meteorite (courtesy Hans Pflug), C is a bacterial fossil within a Brownlee particle and D is a clump of cocci and a bacillus from dust collected using a cryoprobe from 41km in the atmosphere. References and credits are given in Wickramasinghe et al., 2010[17].

Figure 7 shows a much more mysterious organism, taken to be extraterrestrial and potentially an important link to the origin of life. It is the Red Rain material that fell on Kerala, India, in 2001, soon after a sonic boom announced the arrival of a meteor[18].

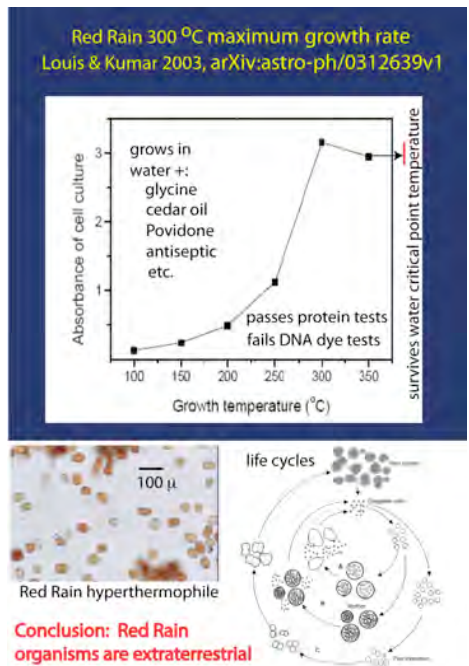

Figure 7. Louis and Kumar study[18] of Kerala Red Rain organisms. Maximum growth rates occur at a record 300°C, far above survival temperatures of known terrestrial hyperthermophiles. Survival temperatures exceed the water critical point 374 °C (red line), which is suggested[19] as optimum for the formation of life chemicals.

The Louis and Kumar (2003)[18] study supports their conclusion that Red Rain organisms are living extraterrestrial hyperthermophiles, delivered to the Earth by cometary panspermia. Complex life cycles (Fig. 7 bottom right) and protein tests suggest DNA regulation. However, the organism fails L-amino acid based DNA dye tests, suggesting Red Rain organisms may be representative of an R-amino acid "shadow biosphere". The handedness of sugars and amino acids are chemically arbitrary, so DNA life should appear in either R or D forms. Figure 8 shows results from other laboratories[20] that generally replicate the Louis and Kumar results.

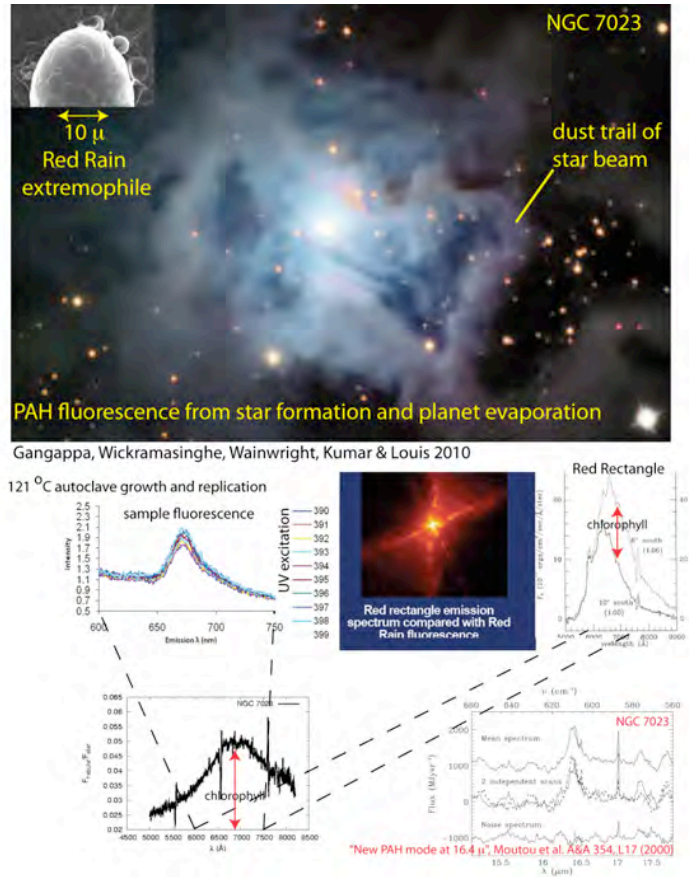

Figure 8. Polycyclic aromatic hydrocarbon PAH evidence of Red Rain organisms in the reflection nebula NGC 7023, the Red Rectangle, and in laboratory replications of the Louis growth results to 121 $^{o}$C. A new PAH mode in the characteristic Red Rain frequency band is reported by Moutou et al. (2000)[21]. Chlorophyll lines appear in both 7023 and the Red Rectangle.

## 5. DISCUSSION

Table 1 and Figures 1 and 2 review the large differences between cosmologies with (HGD) and without (ΛCDMHC) fluid mechanics. HGD cosmology shows life should be primordial, widely distributed and chemically homogeneous. ΛCDMHC cosmology rules out life without miracles. Figures 3, 4 and 5 of the Helix planetary nebula support the HGD conclusion that the dark matter of galaxies is primordial planets in protoglobularstarcluster PGC clumps of a trillion. Figure 6 from the Murchison carbonaceous meteorite shows some of the first strong evidence of cometary panspermia on cosmic scales of extraterrestrial microfossils, as championed for many decades by N. C. Wickramasinge and F. Hoyle.

Figures 7 and 8 describe Red Rain evidence that extraterrestrial life exists with resistance to temperatures and chemicals terrestrial life cannot tolerate, at temperatures exceeding the critical point of water. The critical conditions of water are to astrobiology as the Planck conditions are to the big bang; that is, ideal. Chirality of DNA suggests a binary biosphere on PGC scales.

Figure 9 suggests a likely scenario for first life to occur in the hot gas primordial planets of HGD cosmology. The universe cools as it expands and the primordial planets will also cool by radiative heat transfer to space. Stars and supernovae at protogalaxy cores supply stardust chemicals of life to primordial planet clouds as they cool toward the critical temperature of water 674 K.

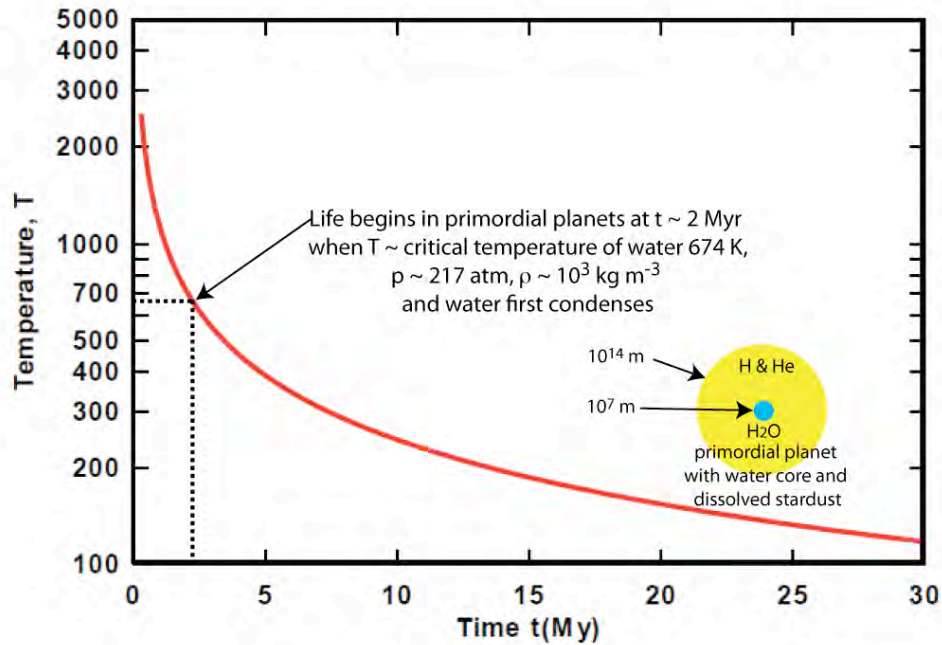

Figure 9. The optimum time and conditions for first life to form in primordial planets is when water first condenses in the presence of the necessary stardust chemicals C, O, N, P etc. following the first stars and supernovae. This occurs at approximately 2 million years after the big bang event[22]. The density of the universe at that time was a million times larger than it is at present.

In this highly reducing atmosphere, iron and nickel from SNe II events collect at planet cores and melt from mergers. Stratified layers of liquids and gases form above the core, including water. From Fig. 9, critical temperature water should start to condense and form life chemicals, and life, at the core of the planets (with dissolved life chemicals C, O, N, P, Si, etc.) at about 2 million years.

Figure 10 shows the present scenario for the formation of life in the first oceans of the communicating primordial planets. The oceans condense at the critical temperature of water 647 K. Critical temperature water is apolar because of the existence of dimers and trimers[19]. The high pressure hydrogen produces mostly apolar molecules like carbon monoxide, methane, and hydrocyanic acid that will readily dissolve and rapidly form more complex organic molecules at such high temperatures.

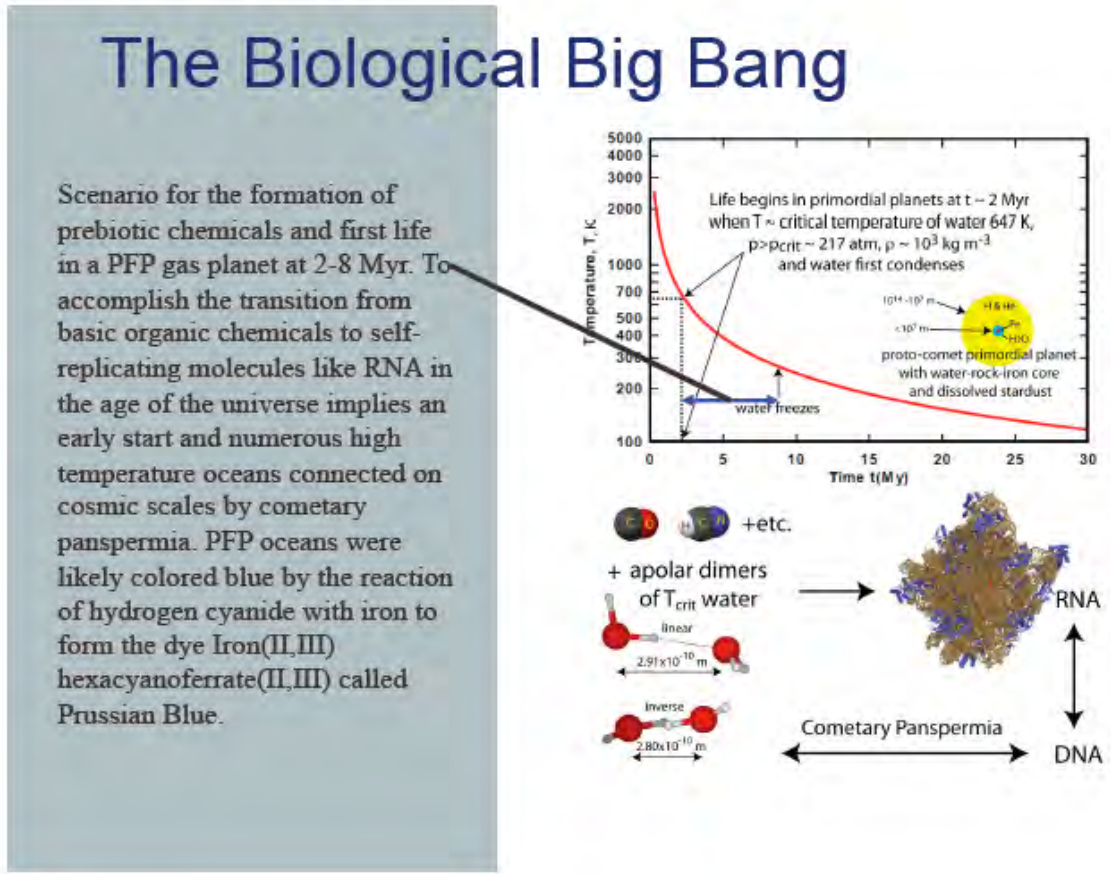

Figure 10. Organic chemistry should evolve rapidly in the first hot water oceans[23]. Self replicating autocatalytic reactions such as RNA have an excellent opportunity to optimize their efficiency in the ~$10^{80}$ hot water oceans produced by the cosmological big bang because all stars are formed by mergers of the primordial planets[24].

Figure 11 shows further evidence of cometary panspermia in the Helix Planetary Nebula from the infrared sensitive satellite Spitzer. In the optical frequency band, no comets can be seen within the Oort cavity, and these are required to explain why the central white dwarf continues to grow until it reaches the Chandrasekhar limit mass of 1.44 solar masses of carbon and explodes to form a supernova 1a. In the infrared, such comets are clearly seen.

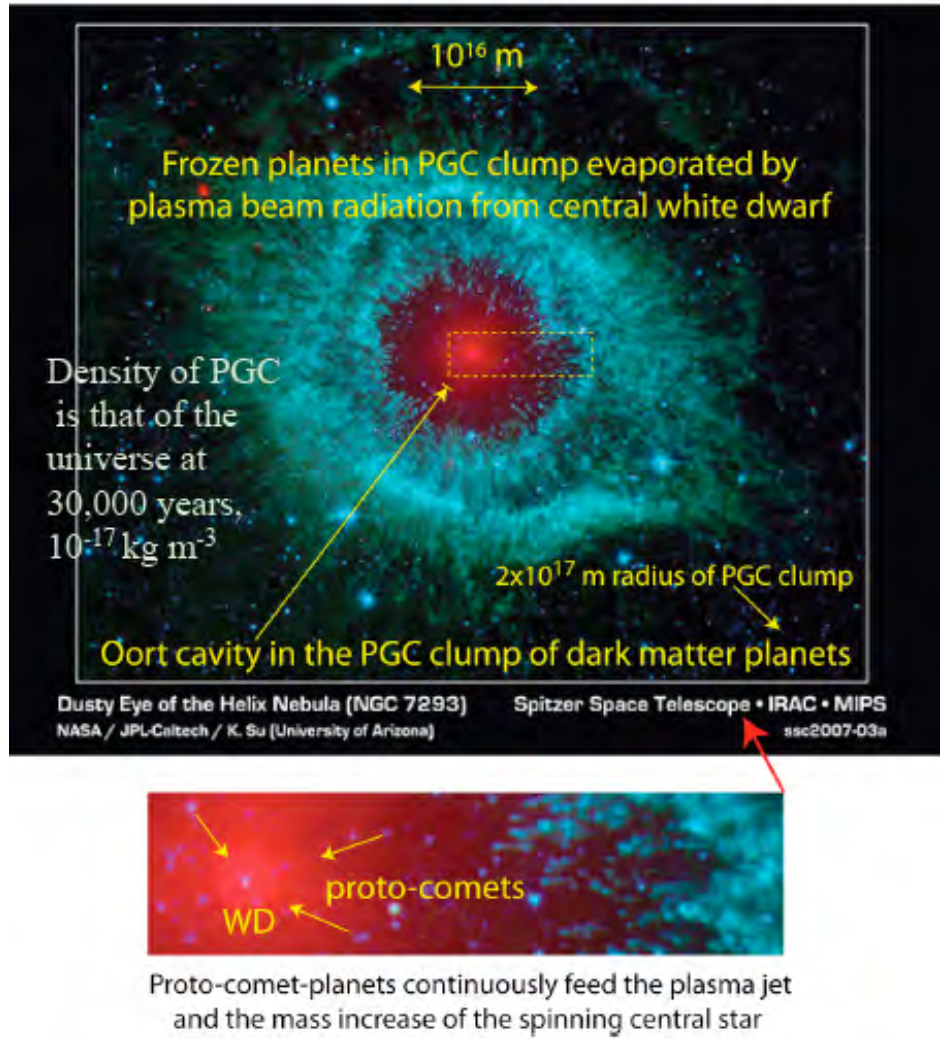

Figure 11. Helix planetary nebula in the infrared from the Spitzer Space Telescope. The central region devoid of planets is termed the Oort cavity. The size of the cavity reflects the mass of one or two stars formed by planet mergers within the PGC clump of planets assuming the density is that of the universe at the time $10^{12}$ seconds when structure first began to form.

Figure 12 is one of the first results of the recently commissioned Herschell Space Observatory, showing unexpected jets of water emerging from young stars in star forming regions of the Milky Way Galaxy. Both the Planck satellite and the Herschell Space Observatory were placed at the second Lagrange point of the Earth-Sun system in May 2009. The water jets are consistent with the HGD cosmology scenario where stars form from merging primordial planets, some of which have frozen water oceans, and unexpected by ΛCDMHD where stars are supposed to form from dry interstellar gas and dust.

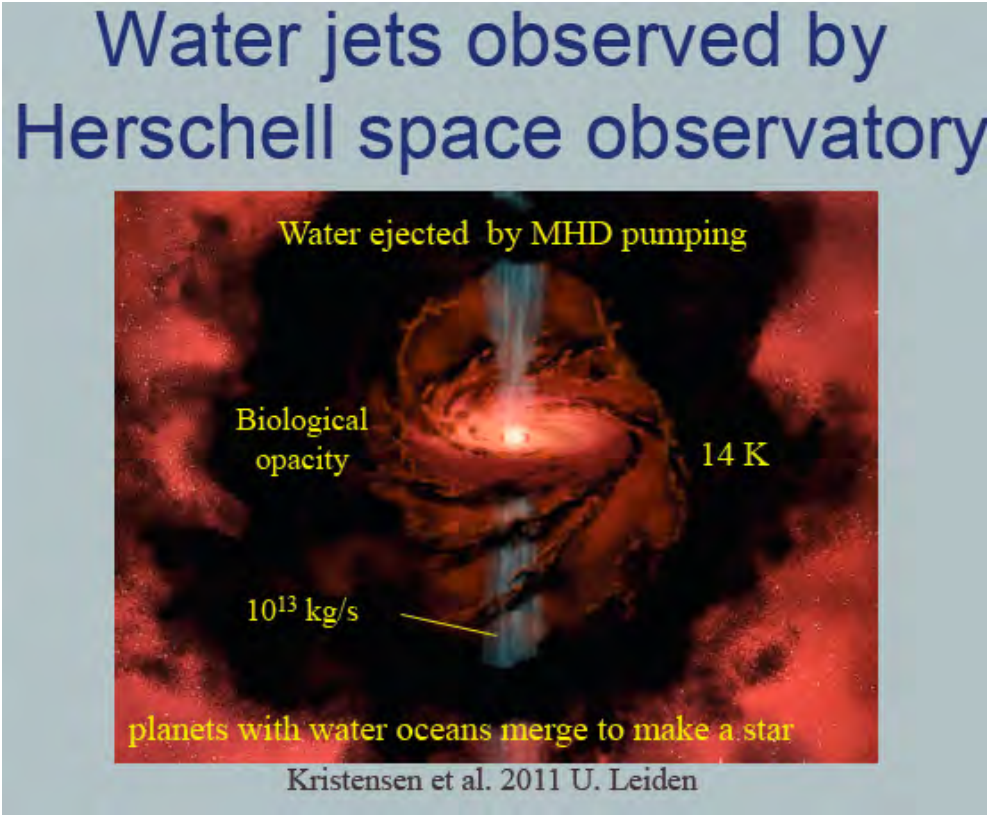

Figure 12. Water jets of $10^{13}$ kg s$^{-1}$ are observed in the first Herschell Space Observatory results. Far infrared images of the jets would normally be obscured by PAH dust in the optical. The "dust" appears to be frozen hydrogen planets from their temperatures of 14 K matching the triple point temperature of hydrogen.

Lars Kristensen of Leiden University compares water flow rates from young stars in terms of flow rates of the Amazon river in a recent National Geographic article.

Figure 13 is an HGD modification of the NASA, WMAP 2006, time line for the Concordance Cosmology CC scenario for the evolution of the universe. Both cosmologies assume a big bang beginning at time zero. On the right, dark energy Λ drives an accelerating expansion of the universe according to CC cosmology, but a big crunch is expected for HGD since entropy is produced by the turbulent big bang mechanism.

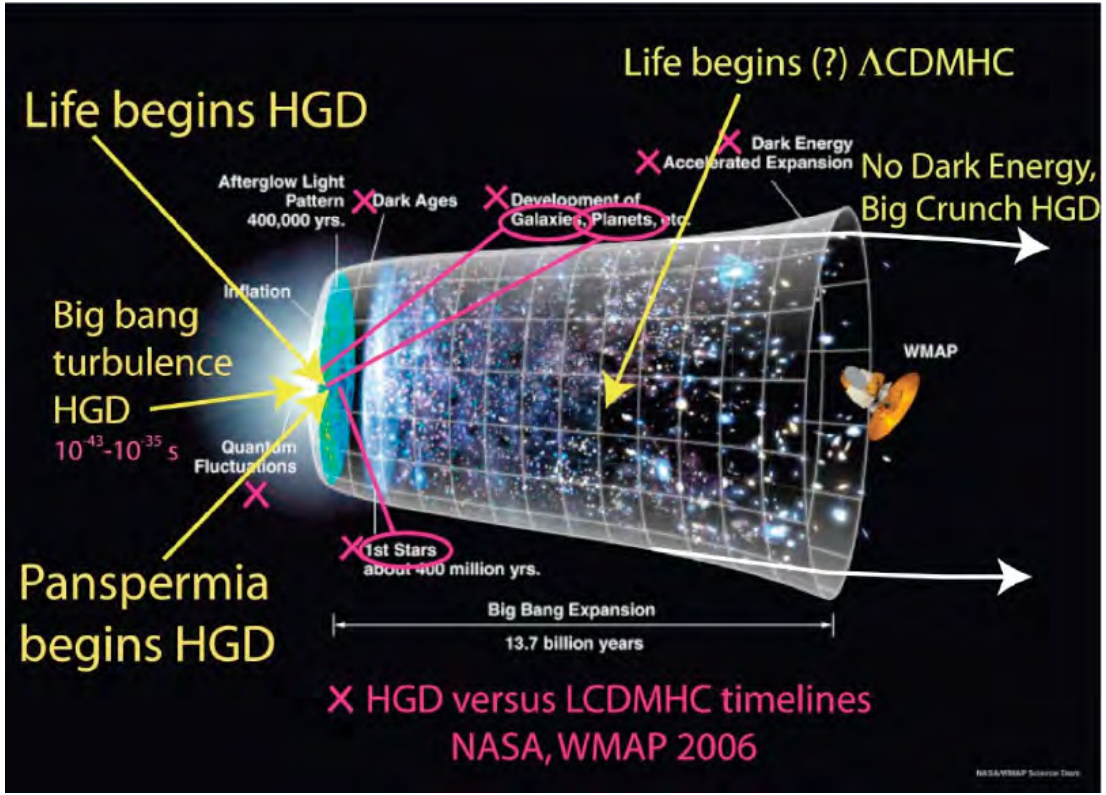

Figure 13. Comparison of time lines for HGD cosmology and ΛCDMHC cosmology. Life formation is virtually impossible in the ΛCDMHC cosmology because stars do not form for 400 million years, with only a handful of planets per star. Life formation in HGD cosmology is early and inevitable.

Figure 14 shows evidence of the relative ages of the RNA subunits in the rRNA 23S complex molecule described by George E. Fox et al. in paper 8152-32 of this meeting[25]. Using modern instrumentation it is possible to identify relative ages of different components of the molecule and its many subunits, looking for the Last Universal Common Ancestor (LUCA) in the phylogenetic tree of life.

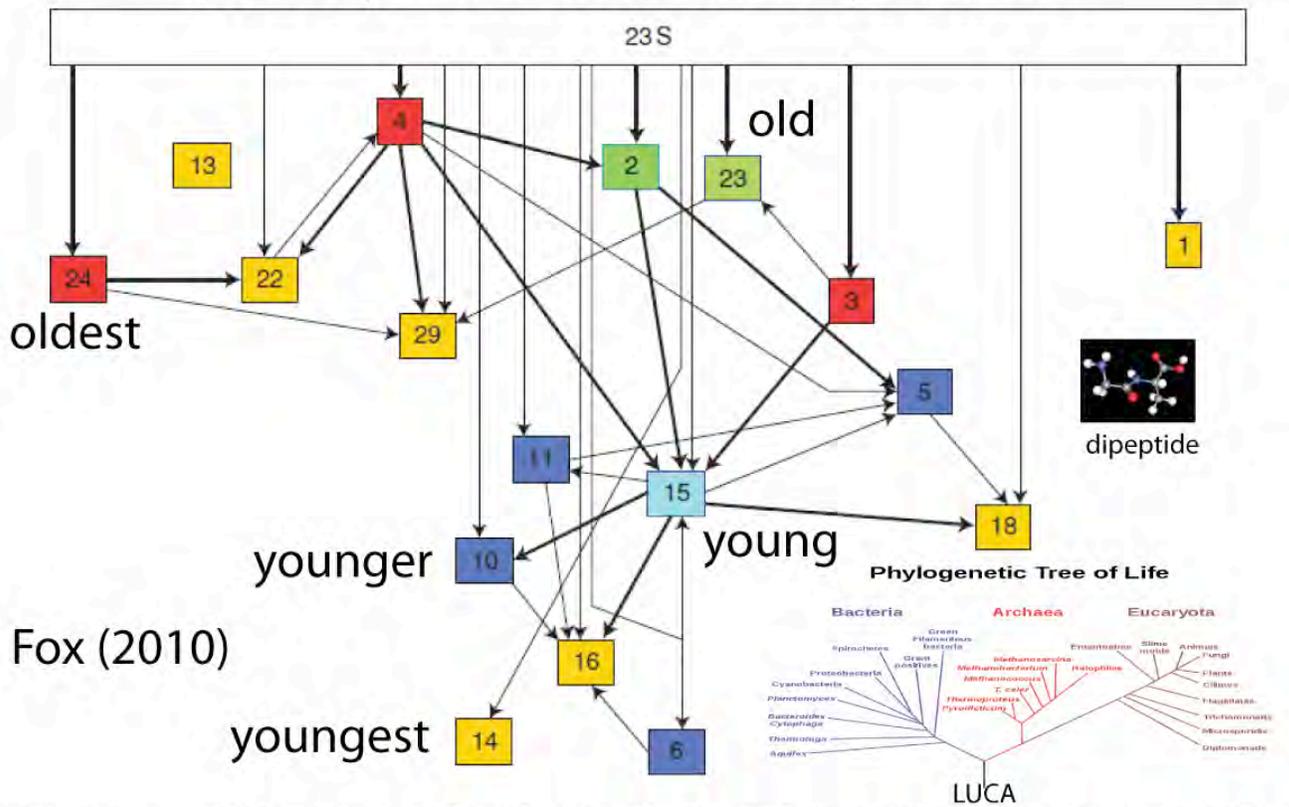

Figure 14. Assembly of RNA subunit 50S by the ribosome complex rRNA 23S. The youngest part is at the center, called the peptidyl transferase center PTC. The PTC appears to be the oldest part of the system, and has no proteins (polypeptides) of its own.

Figure 15 is from one of Loren D. Williams papers[26] mentioned in his contribution 8152-33 of the present symposium. Ribosomes are discussed as "ancient molecular fossils".

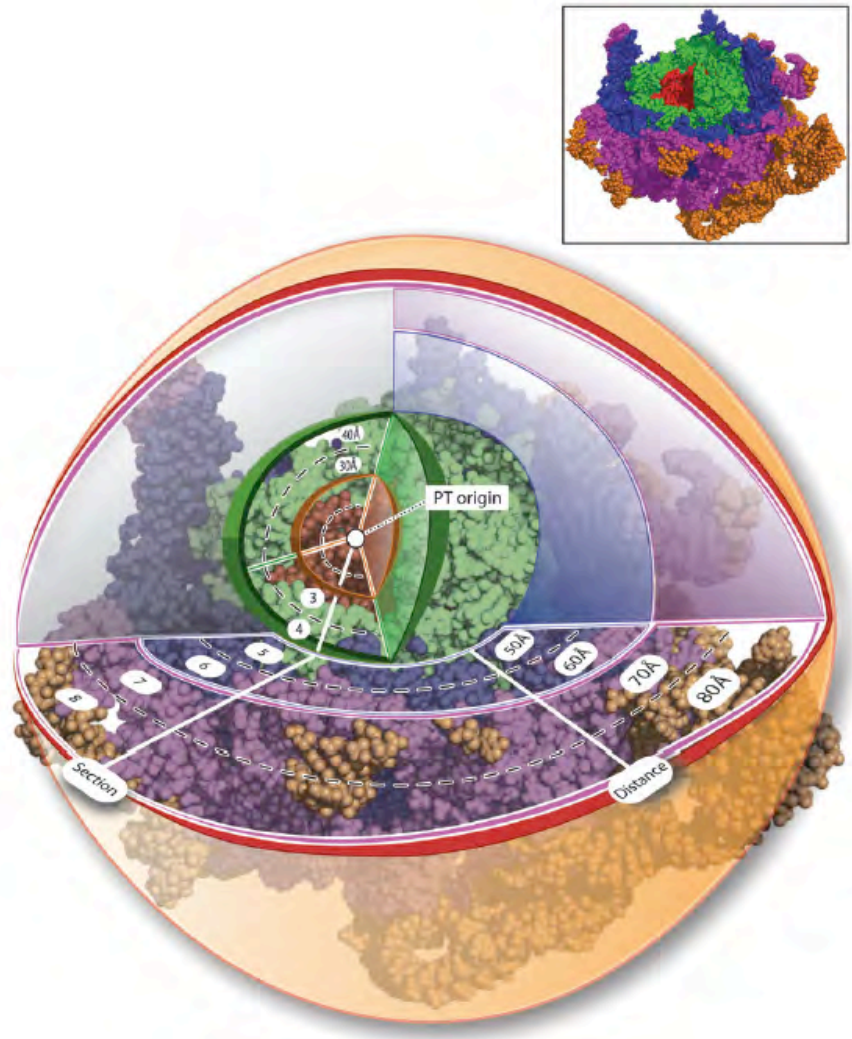

Figure 15. The rRNA 23S ribosome molecule has an onion like structure, with the oldest part of the onion at the center.

The PT-origin in Fig. 15 is the same as the PTC peptidyl transferase center of Fox et al. At this time only relative ages of the rDNA 23S components are proposed. From HGD cosmology, the absolute age of RNA and DNA life is most likely primordial. Nothing from our experience of evolution on Earth rules out primordial intelligent life.

## 6. CONCLUSIONS

Theory and observational evidence show that the standard dark-energy cold-dark-matter hierarchical-clustering ΛCDMHC model for cosmology must be replaced by hydrogravitational dynamics cosmology HGD. An important result of HGD cosmology is its prediction that the dark matter of galaxies is Primordial-Fog-Particle planets PFPs in Proto-Globular-Cluster clumps PGCs. All stars form in these PGC clumps as PFPs merge. The optimum time for first life and the spreading of the seeds of life is very early while the planets are warm and the universe is dense.

We suggest[24] the time of first life was at ~ 2 Myr when the universe cooled sufficiently for water to condense, with stardust fertilizer C, N, O, P etc., at the core of the H-He$^4$ primordial planet clouds, Fig. 9. The critical temperature of water 674 K approximately matches the breakdown temperature of amino acids needed for DNA, and permits the high speed chemical kinetics necessary for living organisms and their complex chemicals to rapidly develop and evolve the highly efficient DNA mechanisms we see on Earth. Chlorophyll catalysts for converting carbon to food can resist such high temperatures. A shadow biosphere with reversed DNA chirality is suggested by studies of the Red Rain organism, whose DNA is undetected by standard methods but whose life cycles and astrophysical signatures imply DNA capabilities. Physical and biological evidence in Figs. 10-15 support a biological big bang description of primordial life formation.

Part of the astrobiological processes that produce life in primordial planets is the distribution of the templates or seeds of life by the plasma jets of the stars the planets produce, as well as the jets from active galactic nuclei that devour millions of stars and eject to other galaxies trillions of life infested planets and planet clumps. The biosphere and shadow biosphere therefore extend to all material produced by the big bang, about $10^{80}$ planets. Biological processes are extremely efficient at converting the carbon of planets to organic chemicals and their fossils, as we see on Earth. With high probability, life did not begin on Earth, or on any of the $10^{18}$ planets of the Galaxy, but was more likely brought to Earth and the Milky Way by cometary panspermia. Biology and medicine are thus subsets of astrobiology on cosmic scales yet to be determined by future studies.